\documentclass[graybox]{svmult}

\usepackage{mathptmx}       
\usepackage{helvet}         
\usepackage{courier}        
\usepackage{type1cm}        
\usepackage{amsmath}
\usepackage{amssymb}
\usepackage{makeidx}         
\usepackage{graphicx}        
\usepackage{multicol}        
\usepackage[bottom]{footmisc}

\newcommand{\doverline}[1]{\overline{\overline{#1}}}

\makeindex             

\begin{document}

\title*{Clustering of functional boxplots for multiple streaming time series}
\author{Elvira Romano and Antonio Balzanella}
\institute{Elvira Romano \at Department of European and Mediterrean Studies, Second University of Naples, Caserta, Italy, \email{elvira.romano@unina2.it}
\and Antonio Balzanella \at Department of European and Mediterrean Studies, Second University of Naples, Caserta, Italy \email{antonio.balzanella@unina2.it}}
%
%
\maketitle

\abstract*{Each chapter should be preceded by an abstract (10--15 lines long) that summarizes the content. The abstract will appear \textit{online} at \url{www.SpringerLink.com} and be available with unrestricted access. This allows unregistered users to read the abstract as a teaser for the complete chapter. As a general rule the abstracts will not appear in the printed version of your book unless it is the style of your particular book or that of the series to which your book belongs.
Please use the 'starred' version of the new Springer \texttt{abstract} command for typesetting the text of the online abstracts (cf. source file of this chapter template \texttt{abstract}) and include them with the source files of your manuscript. Use the plain \texttt{abstract} command if the abstract is also to appear in the printed version of the book.}

\abstract{ In this paper we introduce a micro-clustering strategy for Functional Boxplots. The aim is to summarize a set of streaming time series splitted  in non overlapping windows.
It is a two step strategy which  performs at first, an on-line summarization by means of functional data structures, named Functional Boxplot micro-clusters; then it reveals the final summarization by processing, off-line, the functional data structures.  Our main contribute consists in providing a new definition of micro-cluster based on Functional Boxplots and, in defining a  proximity measure which allows to compare and update them. This allows to get a finer graphical summarization of the streaming time series by five functional basic statistics of data. The obtained synthesis will be able to keep track of the dynamic evolution of the multiple streams.}

\keywords{Time series data stream, Clustream, Micro-clustering, Functional Boxplot}

\section{Introduction}
\label{sec:1}
Data stream mining has gained a lot of attention due to the development of applications where sensor networks are used for monitoring physical quantities such as electricity consumptions, environmental variables, computer network traffic.
In these applications it is necessary to analyze potentially infinite flows of temporally ordered observations which cannot be stored and which have to be processed using reduced computational resources. The on-line nature of these data streams require the development of incremental learning methods which update the knowledge about the monitored phenomenon every time a new observation is collected.

Among the exploratory tools for data stream processing, clustering methods are widely used knowledge extraction tools. Clustering methods in this framework, are used to deal with two problems. The first is to identify, from a set of data streams, groups of streams having similar behavior. This is usually known as clustering of time series data streams \cite{Gama} and some of the main proposals are  \cite{Balzanella}\cite{Ber}\cite{cod}. The second problem is to group the observations that compose a data stream or a set of data streams into homogeneous clusters \cite{Guha}, \cite{Agg}. In this case, the observations available at any given moment, for the different streams, constitute a p-dimensional (where p is the number of data streams) data point. Thus the aim of the on-line algorithm will be to find homogeneous groups of the recorded p-dimensional data points.

Usually these methods also perform the task of summarizing the observed data. This is accomplished by identifying a set of centroids which provides the synthesis of each homogeneous group of observations.
Since the on-line arriving observations are deleted after being processed, the type of adopted synthesis is a key point in the development of methodologies for data stream clustering.

Following this second type of approaches, the CluStream algorithm proposed in \cite{Agg}, provides a two-step strategy. The first is an on-line step, named micro-clustering, that performs a first on-line summarization of the streams keeping updated a specific set of data structures (micro-clusters). The second, is an off-line step named macro-clustering, which reveals the final summarization by processing the micro-clusters with an appropriate clustering algorithm.
The CluStream provides only a basic summarization of the data coming from sensors since it only records the average and the variance of groups of similar multidimensional items. In this paper we extend this algorithm in order to use the Functional Boxplot introduced in \cite{Sun} as tool for gaining knowledge from multiple streaming time series. This will allow to get a finer summarization of the streaming time series that keeps into account five basic statistics (first and third quartile, median, maximum and minimum value) of data and which can be graphically represented.

\section{CluStream of Functional Boxplots }
Let ${y_i}(t),\ \ i=1, \ldots, n,\,t\in [1,\infty]$ a set of streaming time series made by real valued ordered observations of a variable $Y(t)$ in $n$ sites, on a discrete time grid.
This work proposes an incremental clustering algorithm with the aim to supply a set of data descriptions or synopsis to reduce
dimensionality and to keep track of the dynamic evolution of the streams. It processes each example in constant time and memory and is incremental in in the sense that data synopsis are incrementally maintained as more and more data are received.

It is a Clustream algorithm on Functional boxplots obtained by a set of $n$ streaming time series split in non overlapping windows and opportunely approximated by functional data. The method can be summarized by the following steps:

\begin{itemize}
	\item On-line phase (FBP-micro-clustering)
	\begin{itemize}
	    \item Splitting of incoming data streams into non overlapping windows;
	    \item Detection of the Functional Boxplot associated to each window;
	    \item Updating of appropriate data synopsis called Functional Boxplot micro-clusters.
  \end{itemize}
	
	\item Off-line phase (FBP-macro-clustering)
\begin{itemize}
	    \item Clustering algorithm performed on the Functional Boxplot micro-clusters.
\end{itemize}
\end{itemize}

\subsection{On-line phase}

The first step of the on-line phase, consists in splitting the incoming parallel streaming time series into a set of non overlapping windows $W_j, j=1,\ldots,\infty$, that are compact subsets of $T$ having size $w\in \Re$ and such that $W_{j}\bigcap W_{j+1}=\emptyset$. The defined windows frame for each $y_{i}(t)$ a subset $y_{i}^{w_j}(t)\ \ t\in W_{j}$ of ordered values of $y_{i}(t)$, called subsequence.

Following the Functional Data Analysis approach \cite{Ramsay},  we consider each subsequence ${y_{i}}^{w_j}(t)$ of $y_{i}(t)$ the raw data which includes noise information.
Then we determinate a true functional form ${f_{i}^{w_j}}(t)$, we call functional subsequence, which describes the
trend of the flowing data.
For each $W_{j}$ we have that all the subsequences ${y_{i}^{w_j}(t)}\ \ i=1,\ldots,n$ follow the model:
\begin {equation}
y_{i}^{w_j}(t)={f_{i}^{w_j}(t)}+{\epsilon_{i}^{w_j}(t)},\ t\in W_j \ \ i=1,\ldots,n
\end{equation}
where ${\epsilon_{i}^{w_j}(t)}$ are residuals with independent zero mean and $f_{i}^{w_j}(\cdot)$ is the mean function.
\\

The second step of the on-line phase aims at detecting a summary of the set ${f_{i}^{w_j}}(t)$ (with $i=1,\ldots,n$) of the batched streaming time series by means of a functional boxplot variables $FBP_{j}, j=1,\ldots,\infty$, defined as follows:
\begin{definition}[Functional Boxplot]
Let $W_j$ be a window which frames the subsequnces ${f_{1}}^{w_j}(t),\ldots,{f_{i}}^{w_j}(t),\ldots,{f_{n}}^{w_j}(t)$ (with $t\in W_{j}$). A Functional Boxplot $FBP_{j}$ is a compound of five functions $\left\{f_{[u]}^{w_{j}}(t), f_{[l]}^{w_{j}}(t), f_{[1]}^{w_{j}}(t),f_{[b_{min}]}^{w_{j}}(t), f_{[b_{max}]}^{w_{j}}(t)\right\}$ where:

$f_{[u]}^{w_{j}}(t)$ is the upper bound of the central region;

$f_{[l]}^{w_{j}}(t)$ is the lower bound of the central region;

$f_{[1]}^{w_{j}}(t)$ is the median curve

$f_{[b_{min}]}^{w_{j}}(t)$ is the upper bound of the subsequences

$f_{[b_{max}]}^{w_{j}}(t)$ is the lower bound of the subsequences

\end{definition}

A Functional Boxplot is the analog of classical boxplot for functional data \cite{Sun}. The only difference consists in the data ordering criterion. In particular, since functions varies over a continuum, data ordering is based on the notion of band depth or modified band depth \cite{Lopez}.

Based on the center outward ordering induced by band depth for functional data, the descriptive statistics of a functional boxplot are: the envelope of the $50\%$ central region, the median curve, and the maximum non-outlying envelope.
The $50\%$ central region is the analog to the "interquartile range" (IQR), it is defined by the band delimited by the $50\%$ of deepest, or the most central observations. The border of the $50\%$ central region is defined as the envelope representing the box in a classical boxplot.
The median is the most central observation in the box.
The maximum envelope of the dataset identified by the vertical lines of the plot are the "whiskers" of the boxplot. Formally, let $f_{[i]}^{w_{j}}(t)$ denote the sample of functional subsequence associated to the $i$th largest band depth value. The set $f_{[1]}^{w_{j}}(t)\ldots, f_{[n]}^{w_{j}}(t)$ are order statistics, with $f_{[1]}^{w_{j}}(t)$ the median curve, that is  the most central curve (the deepest), and $f_{[n]}^{w_{j}}(t)$ is the most outlying curve.
The central region of the boxplot is defined as
\begin{equation}
C_{0.5}=\left\{(t,f^{w_{j}}(t)): \min_{r=1, \ldots, [n/2]}f_{[r]}^{w_{j}}(t)\leq f^{w_j}(t) \leq \max_{r=1, \ldots, [n/2]}f_{[r]}^{w_{j}}(t)\right\}
\end{equation}
 where $[n/2]$ is the small integer  not less than $n/2$.



In the third step of the on-line phase, the $FBP_{j}$ variables concur to update a set of specific data structures $FBP_{C_k},\:\:k=1,\ldots, K$ we name FBP-micro-clusters, defined as:

\begin{definition}[Functional Boxplot Microcluster]
A FBP-micro-cluster $FBP_{C_k},\:\:k=1,\ldots, K$, for a set of $FBP_{j}$ (with $j=1,\ldots,n^k$) of functional boxplots is the tuple $(\overline{FBP}_{k}, n^k, tl^k, th)$ where:
\begin{itemize}
  \item $\overline{FBP}_{k}$ is the functional boxplot which assumes the role of centroid;
  \item $n^k$ is the number of allocated functional boxplots;
  \item $tl^k$ is the time stamp of the last update;
  \item $th$ is a boundary value
\end{itemize}

\end{definition}

The Functional Boxplot micro-cluster is an extension of the micro-cluster introduced in \cite{Agg}. In our method, its task is to summarize very similar Functional boxplots, through a set of statistics which are updatable on-line and able to adapt to the change of data.

In order to achieve the desired space saving, we keep a set of $FBP_{C_k}$ with $k=1,\ldots,K$ where $K$ is chosen to keep a high representativity of data. Thus $K$ is much higher than the clusters in data but much lower than the number of processed windows.

In the on-line step, every time the data of a new window $W_j$ become available, a $FBP_{j}$ is constructed and then allocated to a $FBP_{C_k}$. The allocation is obtained evaluating the distance between the $FBP_{j}$ and the centroid  $\overline{FBP}_{k}$ so that if the minimum value of distance is lower than the threshold value $th$ stored in the micro-cluster, the allocation is performed to the corresponding $FBP_{C_k}$, otherwise a new one is started setting the functional boxplot of the window as centroid and $n^k=1$.


The allocation is based on the definition of an appropriate distance measure for comparing $FBP_{j}$. It is computed by considering that each couple of correspondent functions is compared on the same time interval by means of an alignment of the $FBP_{j}$.

Let us consider two functional boxplots $FBP_{j}$, $FBP_{j'}$ defined on two windows $W_{j},W_{j'}$. Each of them is characterized by the set of five functions ${f_{.}}^{W_{j}}(t):{W_{j}}\longrightarrow \Re $, ${f_{.}}^{W_{j'}}(t):{W_{j'}}\longrightarrow \Re $.

Aligning $FBP_{j'}$ to $FBP_{j}$ means finding a function $g(t):W_{j'} \longrightarrow W_{j} $ such that ${f_{.}}^{W_{j}}(t)$ and
$g\circ {f_{.}}^{W_{j'}}(t)={h_{.}}^{W_{j}}(t)$ are defined on the same interval $W_{j}$, with the function $g(t)$ expressed by $g(t)=a+bt$.



We consider $a\in \Re$ and $b=1$, that is an alignment. If $b\neq 1$, that is for not only misaligned but also warped functions, the function $g(t)$ can be considered a warping function as in \cite{Sangalli}\cite{Adelfio}.

Thus, formally,  the distance between a pair of functional boxplots $FBP_{j}$, $FBP_{j'}$ is defined as follows:
\begin{definition}[Distance]
Let $FBP_{j}=\left\{f_{[u]}^{w_{j}}(t), f_{[l]}^{w_{j}}(t), f_{[1]}^{w_{j}}(t),f_{[b_{min}]}^{w_{j}}(t), f_{[b_{max}]}^{w_{j}}(t)\right\}$ and  $FBP_{j'}=\left\{f_{[u]}^{w_{j'}}(t), f_{[l]}^{w_{j'}}(t), f_{[1]}^{w_{j'}}(t),f_{[b_{min}]}^{w_{j'}}(t), f_{[b_{max}]}^{w_{j'}}(t)\right\}$ be two functional boxplots defined, respectively, on $W_j$ and $W_{j'}$, $g(t)$ be the alignment function so that $g\circ {f_{.}}^{W_{j'}}(t)={h_{.}}^{W_{j}}(t)$, the distance between $FBP_{j}$ and $FBP_{j'}$ is:
\begin{eqnarray*}
d(FBP_{j}, FBP_{j'})& = & \sqrt{\int_{t\in W}(f_{[u]}^{w_{j}}(t)-h_{[u]}^{w_{j}}(t))^2 dt} + \sqrt{\int_{t\in W}(f_{[l]}^{w_{j}}(t)-h_{[l]}^{w_{j}}(t))^2 dt}+\\
 & & +\sqrt{\int_{t\in W}(f_{[1]}^{w_{j}}(t)-h_{[1]}^{w_{j}}(t)^2 dt)}+\sqrt{\int_{t\in W}(f_{[b_{min}]}^{w_{j}}(t)-h_{[b_{min}]}^{w_{j}}(t))^2 dt}+\\
 & & +\sqrt{\int_{t\in W}(f_{[b_{max}]}^{w_{j}}(t)-h_{[b_{max}]}^{w_{j}}(t))^2 dt}
\end{eqnarray*}
\label{distanza}
\end{definition}

The consequences of an allocation are the unitary increment of $n^k$, the setting of the current time stamp for the parameter $W^k$ and the computation of the FPB-micro-cluster centroid. The latter is performed  so that for each of the five functions which define the Functional Boxplot, the average is kept. This can be obtained starting from the information stored in the FBP-micro-cluster self and from the just allocated Functional Boxplot.

In our method, the size $K$ of the set of FPB-micro-cluster is not defined a-priori but it adapts to the structure of data, however it strongly depends on the choice of the threshold $th$. A too high value involves that only few FBP-micro-clusters are generated; on the contrary, a too low value brings to generate too many FBP-micro-clusters. To deal with this issue we introduce an heuristic to set the value of the threshold and a criterion to keep the number of functional boxplot micro-clusters under a value $K_{max}$ (this allows to keep a constant upper bound of the used memory space).

Particularly, we propose to compute the threshold $th$ as follows:
\begin{equation}
	th= min \ d(\overline{FBP}_{j},\overline{FBP}_{k}) \ \ \forall j,k=1,\ldots,K \ with \ k\neq j
\end{equation}

thus, $th$ is set to the minimum distance between the FBP-micro-cluster centroids.

If the number of FBP-micro-clusters grows too much so to exceed the available memory resources, we propose, alternatively, to discard the micro-clusters recording concepts no longer present in the data or to merge the two nearest FBP-micro-clusters into one. The choice is made by evaluating the time stamp of the last updating stored in the parameter $tl^k$ of each FBP-micro-cluster:

$$\begin{cases}
\mbox{If} \ \ (t_{now}-tl^k) > t^* \ \ \  \forall k=1,\ldots,K \ \ \ \Rightarrow \ \ Discard \ \mbox{$FBP_{C_k}$}, \ \ \  \\
\mbox{Else} \ \ argmin_{j,k} \ d(\overline{FBP}_{j},\overline{FBP}_{k}) \ \ \forall j,k=1,\ldots,K \ with \ k\neq j  \ \Rightarrow \ \ Merge \ FBP_j,FBP_k
\end{cases}$$

where $t_{now}$ is the time stamp of the current window and $t^*$ indicates the age over which a FBP-micro-cluster has to be considered no longer useful.


\subsection{Off-line phase}
In order to reveal the final summarization of the streams, the off-line phase analyzes the FBP-micro-clusters computed on-line.
We provide a method to get the summarization of data behavior over user defined time slots. It is based on storing, at predefined time instants, a snapshot of the set of FBP-micro-cluster. Each snapshot will collect the state of updating of each $FBP_{C_k}$ in that time instant.

In order to get the summarization of the user defined time slot, the procedure identifies the snapshot that is temporally closer to the lower end of the time interval (lower snaphot) and the one which is temporally closer to the upper end (upper snapshot). The next step is to remove from the state of the functional boxplot micro-clusters the effects of the updates that occurred before the beginning of the lower snapshot.
Since the centroid $\overline{FBP_k}$ of each FBP-micro-cluster is the average of the allocated functional boxplots, it is possible to recover the state of each $FBP_{C_k}$ removing what has happened before the beginning of the time slot, by computing a component by component weighted difference between the centroid $\overline{FBP_k}$ as available from the upper snapshot and the corresponding $\overline{FBP_k}$, obtained from the lower snapshot (the weights are the number of allocations stored in the parameter $n^k$).

From the output of the previous step, the obtained centroids $\overline{FBP_k}$, together with the number of allocated items $n_k$ (which assumes the role of weight), become the data to be processed by a k-means like algorithm which provides, as output, a partition of the FBP-micro-clusters centroids into a set $MC_1, \ldots, MC_c, \ldots, MC_C$ (with $C<K$) of macro-clusters  and a new set $\doverline{FBP_c}$, (with $c=1,\ldots,C$) of functional boxplots which are the final summaries of the required time interval.
Similarly to the k-means, this algorithm minimizes an internal heterogeneity measure:
\begin{equation}
 \Delta= \sum_{c=1}^{C}{\sum_{\overline{FBP_k} \in MC_c} {d(\overline{FBP_k};\doverline{FBP_c})n_k  }}
\end{equation}

where $d(\overline{FBP_k};\doverline{FBP_c})$ is computed according to the definition \ref{distanza}.

In order to optimize the criterion $\Delta$, our macro-clustering algorithm iterates, until the convergence, an allocation and a centroid computation step. In the allocation step, each $\overline{FBP_k}$ is attributed to the macro cluster whose distance is minimal. In the centroid computation step, the representation of each macro-cluster $MC_c$ is obtained by means of a component by component weighted average where $n_k$ is the weight for the corresponding $\overline{FBP_k}$.

\section{Daily Rainfall Monitoring by Clustering of FBP}

\begin{figure}[h]
\includegraphics[scale=.25]{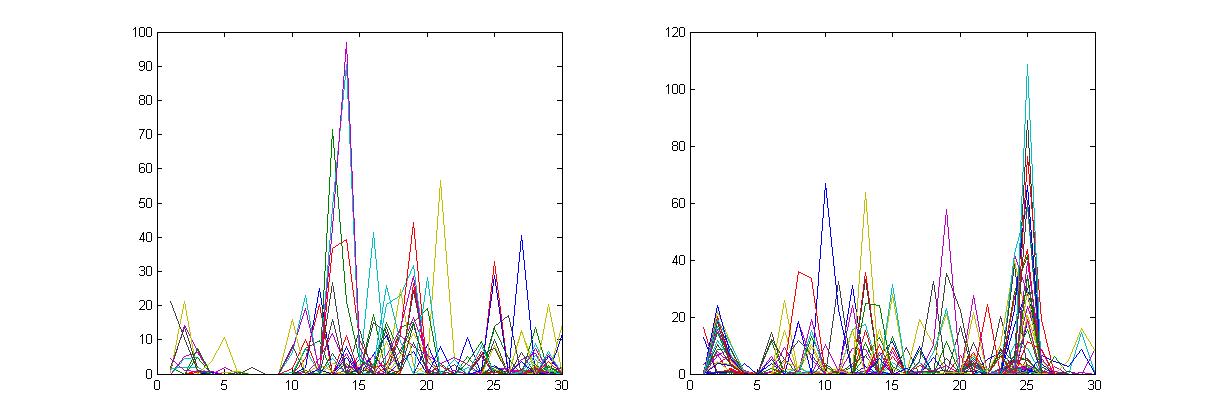}
%
%
\caption{Daily Rainfall in two different time windows made by $30$ observations.}
\label{fig1}       
\end{figure}

This section shows the results on real data of the proposed method. We have analyzed a dataset provided by the Australian Government - Bureau of meteorology, available on-line at $http://www.bom.gov.au/climate/data/$, which records the daily rainfall in Australia from $1/4/1961$ to $30/4/2012$.
We have downloaded $77$ time series, each one made by $15139$ observations and corresponding to a weather station located in the Australia region. The choice of the observation period and the selection of the weather stations has been carried out in such a way to have no missing data.
Precipitation is most often rain, but also includes other forms such as snow. Observations of daily rainfall are nominally made at 9 am local clock time and record the total precipitation for the preceding $24$ hours. If, for some reason, an observation is unable to be made, the next observation is recorded as an accumulation, since the rainfall has been accumulating in the rain gauge since the last reading.
As can be seen from Fig.\ref{fig1}, daily rainfall is characterized by intense variations. The highest values of the mean precipitation
reached in the fifteen days of the first window could seem comparable with the maxima of the twenty five days of the the second window. However it is not the same daily rainfall stream but a stream related to different stations. In this sense, the overall trend of the phenomenon cannot be detected. In the following we show as our method can help to catch the main rainfall behaviors along the whole observation period and to describe and graphically represent them by means of a set of Functional boxplots.

The assessment of the method requires to set two input parameters: the size of each window $w$ and the maximum number $K_{max}$ of generated $FBP_{C_k}$ micro-clusters. We set the first one to $w=30$ in order to get on-line computed functional boxplots summarizing thirty days of observations. The second parameter has been set to $K_{max}=50$, which represents a good compromise between the detail of summarization and the memory usage.

\begin{table*}[htbp]
	\centering

\begin{tabular}{p{1.5cm}p{0.5cm}p{1.5cm}p{0.5cm}p{1.5cm}p{0.5cm}p{1.5cm}p{0.5cm}p{1.5cm}p{0.5cm}}
\hline\noalign{\smallskip}
$FBP_{C_k}$& $n^k$ & $FBP_{C_k}$ & $n^k$ & $FBP_{C_k}$& $n^k$ & $FFBP_{C_k}$& $n^k$ & $FBP_{C_k}$& $n^k$\\
\noalign{\smallskip}\svhline\noalign{\smallskip}
$FBP_{C_1}$ & $45$ & $FBP_{C_{11}}$ & $1$ & $FBP_{C_{21}}$& $1$ & $FBP_{C_{31}}$ & $1$ & $FBP_{C_{41}}$ & $15$\\
$FBP_{C_2}$ & $2$ & $FBP_{C_{12}}$& $1$ & $FBP_{C_{22}}$ & $1$ & $FBP_{C_{32}}$ & $1$ & $FBP_{C_{42}}$ & $1$\\
$FBP_{C_3}$ & $2$ & $FBP_{C_{13}}$ & $4$ & $FBP_{C_{23}}$ & $1$ & $FBP_{C_{33}}$ & $4$ & $FBP_{C_{43}}$ & $1$\\
$FBP_{C_4}$ & $380$ &$FBP_{C_{14}}$ & $4$ & $FBP_{C_{24}}$ & $4$ & $FBP_{C_{34}}$ & $1$ & $FBP_{C_{44}}$& $1$\\
$FBP_{C_5}$ & $26$ & $FBP_{C_{15}}$ & $1$ & $FBP_{C_{25}}$ & $1$ & $FBP_{C_{35}}$ & $1$ & $FBP_{C_{45}}$ & $1$\\
$FBP_{C_6}$ & $1$ & $FBP_{C_{16}}$& $1$ & $FBP_{C_{26}}$ & $1$ & $FBP_{C_{36}}$ & $28$ & $FBP_{C_{46}}$ & $1$\\
$FBP_{C_7}$ & $5$ & $FBP_{C_{17}}$ & $2$ & $FBP_{C_{27}}$ & $1$ & $FBP_{C_{37}}$ & $1$ & $FBP_{C_{47}}$ & $2$\\
$FBP_{C_8}$ & $24$ & $FBP_{C_{18}}$ & $45$ & $FBP_{C_{28}}$ & $2$ &$FBP_{C_{38}}$& $38$ & $FBP_{C_{48}}$ & $1$\\
$FBP_{C_9}$ & $2$ & $FBP_{C_{19}}$ & $4$ & $FBP_{C_{29}}$ & $2$ & $FBP_{C_{39}}$ & $1$ & $FBP_{C_{49}}$& $1$\\
$FBP_{C_{10}}$ & $4$ & $FBP_{C_{20}}$ & $1$ & $FBP_{C_{30}}$ & $1$ & $FBP_{C_{40}}$ & $1$ & $FBP_{C_{50}}$ & $1$\\

\noalign{\smallskip}\hline\noalign{\smallskip}
	
		\end{tabular}
	\caption{The number of on-line computed $FBP$ allocated to each $FBP_{C_k}$}
	\label{tab:ExternalValidityIndices}
\end{table*}

From the results, see Tab.\ref{tab:ExternalValidityIndices}, we can observe that seven $FBP_{C_k}$ collect more than $5$ on-line computed functional boxplots so these are the the ones that record the main concepts in the data. The remaining FBP-micro-clusters summarize the anomalous or residual rainfall behaviors.

\begin{figure}[h]
\includegraphics[scale=.25]{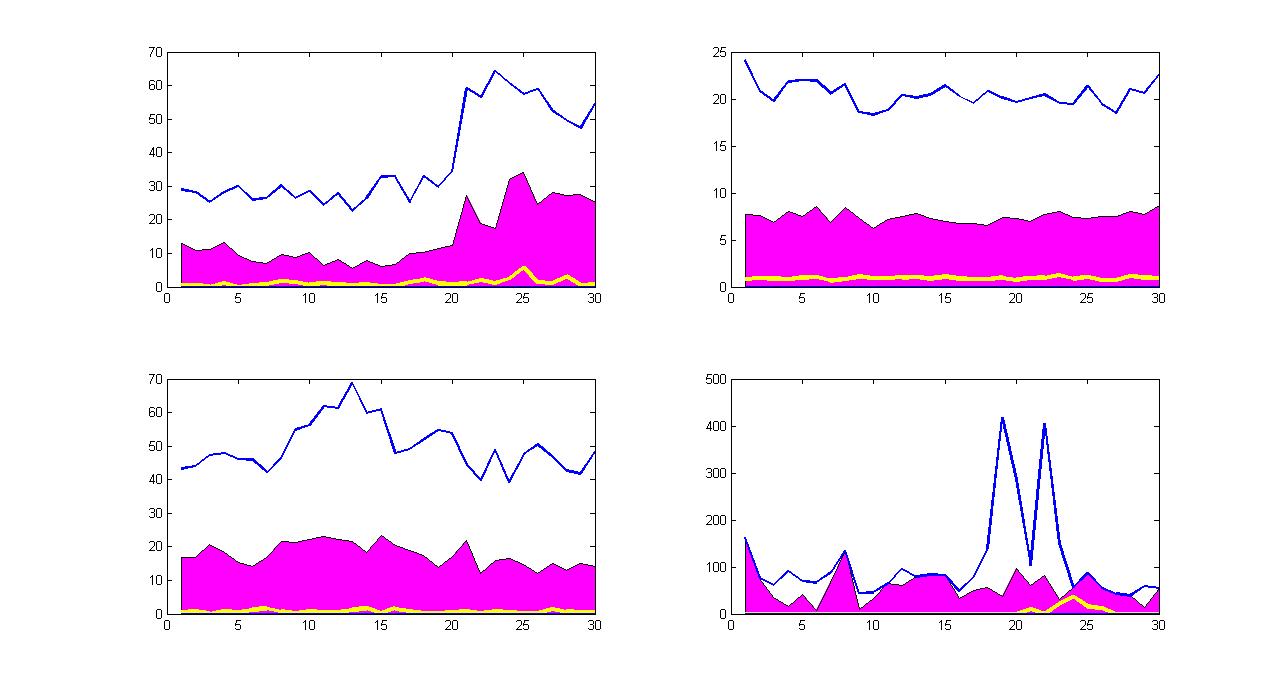}
%
%
\caption{Functional Boxplots summarization for Daily Rainfall Data with four Microcluster.
The blue curves denotes  the envelope. The magenta area delimits the $50\%$ central region. The yellow curve represents the median curve.}
\label{fig2}       
\end{figure}

The off-line procedure, which is performed taking as input the whole set of $FBP_{C_k}$, provides a final summarization of the data. We are interested in discovering how the whole trend changes over the days and if there are dominant structure in the data behaviors. Thus, we choose to get four final functional boxplots summarization( Fig. \ref{fig2}).

Comparing the original curves to the four functional boxplots, we see that the latter are very informative to underline the main changes in the data.
In all the four cases, the curve distributions are asymmetric and positively skewed.
The four functional boxplots differs mostly for: the median curve, that can be interpreted as the most representative observed patterns of rainfall data; the central region, that gives a less biased visualization of the curves' spread.

In the first Functional Boxplot, the median curve is characterized by low and oscillating rainfall trend around $1.5\textit{mm}$ with higher values between the $23\textit{th}$ and $27\textit{th}$ day. In this case, more information is detected by observing the box. It highlights that the $23\textit{rd}$ and $25\textit{th}$ rainfall are high in the last $10$ days.

At the opposite the second Boxplot depicts a quasi constant rainfall trend around the $3\textit{mm}$ (the median curve) with a similar shape of the box but with higher values of the rainfall. This indicates that the trend rainfall vary with a constant trend among $5\textit{mm}$ and $7.5\textit{mm}$.
The third Functional Boxplot instead, shows lower values of the rainfall median curve with an highest values of the box bounds (the values vary among $18mm$ and $20mm$).
Finally the forth Functional Boxplot highlights a median rainfall near to zero except for the $20\textit{th}$ and $28\textit{th}$ and a box with a concentration of rainfall curves in the third quartile with high variability.

All the four Functional Boxplot have an envelope bounded by the blue curve which has a minimum value corresponding to an absence of rainfall. Thus, the lower curve shall be the same with the x-axis. The upper curve limit, indicating the maximum value of the fall of rain, is characterized by four different behaviors linked evidently to different period of summarization. In the first Functional Boxplot it can be observed a curve with an almost constant  trend with a value oscillating around $30\textit{mm}$  for the first twenty days and up to $80\textit{mm}$ in the other $10$ days. In the second and third boxplot on the contrary, the trends vary  around the value of $20\textit{mm}$ and $55\textit{mm}$. Finally the fourth boxplot evidence an oscillation value significantly higher $100\textit{mm}$ with a peak of volatily between $20\textit{th}$ and $25\textit{th}$ day.

%

\section{Concluding remarks}
In this paper we have introduced a new Clustream strategy for  multiple streaming time series.
It is based on a two-step process to handle incremental time series.
In a first step (the online step) graphical summarizing structures, named Functional Boxplots,  continuously updated are detected. In the second step (the offline step) a final graphical summarization of the flow data is obtained.

Unlike the existent CluStream strategy in streaming time series literature, we have introduced a tool able also to provide a graphic synthesis.


\begin{thebibliography}{99.}%
%

%
\bibitem{Adelfio} Adelfio G., Chiodi M., D�Alessandro A., Luzio D., D�Anna G., Mangano G. Simultaneous seismic wave clustering and registration. Computers Geosciences 44, 60�69. ISSN: 0098-3004. DOI: 10.1016/j.cageo.2012.02.017. (2012)
\bibitem{Agg}Aggarwal C. C., Han J., Wang J., Yu P. S. A Framework for Clustering Evolving Data Stream. In Proc. of the 29th VLDB Conference.(2003)
\bibitem{Balzanella} Balzanella A., Lechevallier Y., Verde R. Clustering Multiple Data Streams. In New Perspectives in Statistical Modeling and Data Analysis. Springer. ISBN: 978-3-642-11362-8. DOI: 10.1007/978-3-642-11363-5-28. (2011)
\bibitem{Ber}Beringer J., Hullermeier E. Online clustering of parallel data streams. Data and Knowledge Engineering, 58(2). (2006)
\bibitem{cod} Bi-Ru Dai, Jen-Wei Huang, Mi-Yen Yeh, and Ming-Syan Chen. Adaptive Clustering for Multiple Evolving Streams. In IEEE Transactions On Knowledge And Data Engineering, Vol. 18, No. 9. (2006)


\bibitem{Gama} Gama J., Gaber, M.M (Eds). Learning from Data Streams: Processing Techniques in Sensor Networks.Ed. Springer Verlag. (2007)
\bibitem{Guha} Guha S.,  Meyerson A., Mishra N. and Motwani R. Clustering Data Streams: Theory and practice. IEEE Transactions on Knowledge and
Data Engineering, vol. 15, no. 3, pp. 515-528. (2003)

\bibitem{Lopez} Lopez-Pintado S., Romo J. On the Concept of Depth for Functional Data. Journal of the American
Statistical Association,  \textbf{104}, 718--734, (2009).

\bibitem{Ramsay} Ramsay J.E., Silverman B.W. Functional Data Analysis (Second ed.).Springer. (2005)

\bibitem{Romano}Romano E., Balzanella A., Rivoli L. Functional boxplots for summarizing and detecting changes in environmental data coming from sensors. In Electronic Proceedings of Spatial 2, Spatial Data Methods for Environmental and Ecological Processes 2nd Edition. Foggia, 1-3 Settembre 2011. .
\bibitem{Sangalli} Sangalli L.M., Secchi P., Vantini S.,  Vitelli V. k-mean alignment for curve clustering, Computational Statistics and Data Analysis, Volume 54, Issue 5, 1 May 2010, Pages 1219-1233, ISSN 0167-9473, 10.1016/j.csda.2009.12.008.
    
\bibitem{Sun}Sun Y., Genton M.G.: Functional boxplots. Journal of Computational and Graphical Statistics, \textbf{20}, 316-334. (2011).


\end{thebibliography}
\end{document}